\documentclass[twocolumn,
pra,
 showpacs,floats,preprintnumbers,amsmath,amssymb]{revtex4}
\usepackage{epsfig}
\usepackage{amsmath}
\renewcommand{\(}{\left(}
\renewcommand{\)}{\right )}

\renewcommand{\]}{\right ]}

\def\num{\begin{subequations}}
\def\enum{\end{subequations}}

\def\bea{\arraycolsep .1em \begin{eqnarray}}
\def\eea{\end{eqnarray}}

\let\no=\nonumber

\def\eq#1{Eq.(\ref{#1})}
\def\refr#1{\cite{#1}}
\def\refrs#1{Refs.\cite{#1}}

\def\s0#1#2{\mbox{\small{$ \frac{#1}{#2} $}}}
\def\0#1#2{\frac{#1}{#2}}

\def\anp#1#2#3{Adv.\ Nucl.\ Phys. \ {\bf #1}, #2 (#3)}
\def\cpl#1#2#3{Chin. \ Phys.\ Lett. \ {\bf #1}, #2 (#3)}
\def\ctp#1#2#3{Commun.\ Theor.\ Phys. \ {\bf #1}, #2 (#3)}
\def\plb#1#2#3{Phys. Lett. {\bf B #1}, #2 (#3)}

\def\pra#1#2#3{Phys. Rev.  {\bf A #1}, #2 (#3)}

\def\prc#1#2#3{Phys. Rev.  {\bf C #1}, #2 (#3)}

\def\prl#1#2#3{Phys. Rev. Lett. {\bf #1}, #2 (#3)}
\def\ann#1#2#3{Ann. Phys. (N.Y.) {\bf #1}, #2 (#3)}
\def\anp#1#2#3{Adv. Nucl. Phys. {\bf #1}, #2 (#3)}

\def\epl#1#2#3{Europhys.\ Lett.{\bf #1}, #2 (#3)}
\def\prep#1#2#3{Phys.\ Rep.\ {\bf #1}, #2 (#3)}

\def\njp#1#2#3{New J. Phys.\ {\bf #1}, #2 (#3)}

\def\ijmpe#1#2#3{Int.\ J.\ Mod.\ Phys.\ {\bf E #1}, #2 (#3)}

\def\rmp#1#2#3{Rev.\ Mod.\ Phys.\ {\bf #1}, #2 (#3)}

\def\sci#1#2#3{Science\ {\bf #1}, #2 (#3)}

\begin{document}

\title{The virial equation of state for unitary fermion thermodynamics with non-Gaussian correlations}
\author{Ji-sheng Chen\footnote{chenjs@iopp.ccnu.edu.cn},~~~~~
Jia-rong Li,~~~~~Yan-ping Wang,~~~~~and~~~~~Xiang-jun Xia\\
}

\address{Physics Department \& Institute of Particle Physics, Central China
Normal University, Wuhan 430079, People's Republic of China}

\date{\today}

\begin{abstract}
We study the roles of the dynamical high order perturbation and
statistically non-linear infrared fluctuation/correlation in the
virial equation of state for the Fermi gas in the unitary limit.
Incorporating the quantum level crossing rearrangement effects, the
spontaneously generated entropy departing from the mean-field theory
formalism leads to concise thermodynamical expressions. The
dimensionless virial coefficients with complex non-local
correlations are calculated up to the fourth order for the first
time. The virial coefficients of unitary Fermi gas are found to be
proportional to those of the ideal quantum gas with integer ratios
through a general term formula. Counterintuitively, contrary to
those of the ideal bosons
    ($a^{(0)}_2=-\frac{1}{4 \sqrt{2} }$)
    or fermions($a^{(0)}_2=\frac{1}{4 \sqrt{2} }$),
    the second virial coefficient $a_2$ of
    Fermi gas at
unitarity is found to be equal to zero. With the vanishing leading
order quantum correction, the BCS-BEC crossover thermodynamics
manifests the famous pure classical Boyle's law in the Boltzmann
regime. The non-Gaussian correlation phenomena can be validated by
studying the Joule-Thomson effect.
\end{abstract}

\preprint{Preprint No: 0712.0205} \maketitle

\section{Introduction}
Unconventional unitary fermion physics is associated with a variety
of strongly interacting topics. This theme can test the many-body
theories, from neutron stars and nuclear matter to quark-gluon
plasmas etc.

In recent years, the study for the interacting fermion matter
properties has attracted much attention in quan- tum many-body
community\refr{Giorgini2007}. This is attributed to the rapid
progress of atomic Fermi gas experiments. Controlling the $S$-wave
scattering length between two different spin components allows one
to control the interaction strength by using a magnetically tuned
Feshbach resonance. With the magnetic tuning technique, increasing
the interaction strength of atomic fermions with scattering length
$a$ from $-\infty$ to $+\infty$ resulting in bound boson systems
exhibits the Bardeen-Cooper-Schrieffer to Bose-Einstein
    condensation (BCS-BEC crossover).
The two regimes with positive and negative $S$-wave scattering
    length meet in the strongly interacting limit with the divergent
    scattering length.
At the resonant point, the scattering cross-section will be
    saturated as $\sigma={4\pi}/{k^2}$ (with $k$ being the relative
    wave-vector magnitude of the colliding particles) due to the fundamental unitary
    property limit.
The strongly interacting BCS-BEC crossover topic is literally
    called the unitary Fermi gas thermodynamics\refr{Giorgini2007,Heiselberg2000,Ho2004,Ho2004-1,Thomas2005,Luo2007,Bulgac2005-1,Burovski2006,Schwenk2005,physics/0303094,Pethick,Steele,Xiong2005,Hu2006,Rupak2007,Castin2006}.

Generally, the thermodynamical properties of the low energy dilute
Fermi system are determined by the $S$-wave scattering length $a$,
the particle number density $n$ as well as the temperature $T$. In
the resonant regime with a zero-energy bound state, the divergent
scattering length will certainly drop out in the thermodynamical
quantities; i.e., the thermodynamical properties are
universal\refr{Giorgini2007,Heiselberg2000,Ho2004}. The divergent
scattering length poses an intractable many-body problem.

In addressing the BCS-BEC crossover thermodynamics, the fundamental
    issue is the ground state energy. On the basis of the dimensional analysis,
    the dimensionless coefficient $\xi$ relates the energy per
    particle $E/N= \xi \035 \epsilon _f$ with the Fermi kinetic energy
    $\epsilon_f =k_f^2/(2m)$. Here, $m$ is the bare
    fermion mass while $k_f$ is the Fermi momentum.
The fundamental
    universal coefficient $\xi$ has prompted many theoretical or experimental
    efforts in recent years\refr{Giorgini2007}.
Furthermore, the finite
    temperature thermodynamic properties of unitary Fermi
    gas are as intriguing as the zero-temperature ground state energy and many
    experimental/theoretical efforts have been made\refr{Thomas2005,Luo2007}\refr{Bulgac2005-1,Burovski2006}.

Compared with the zero-temperature ground state energy, the
    additional energy scale, i.e, the reciprocal thermodynamical de
    Broglie wavelength $\lambda^{-1} =\sqrt{{mT}/{(2\pi)}}$ complicates the theme and makes the universal property analysis more
    profound\refr{Ho2004}.
In the weak
    degenerate Boltzmann regime, the universal properties are characterized by the
    virial coefficients \refr{Ho2004,Ho2004-1,Schwenk2005,Rupak2007}.
 For example,
the virial
    equation of state is
    related with the neutrinosphere physics in
    supernovae for the dilute nuclear matter matter.
It is believed that the virial equation of state will influence the
    detailed information of the neutrino response of low-density neutron
    matter\refr{Schwenk2005}.
How to calculate the virial coefficient in the unitary limit
    regime remains an important many-body topic.
Like in addressing the zero-temperature ground state energy problem,
    the central task is to understand the novel non-linear
    fluctuation/correlation physics.

The virial expansion is the basic tool for use in discussing the
    thermodynamical properties and should be model independent.
In thermodynamics, the non-linear virial
    expansion is the \textit{infinite} series of the pressure according to the particle number density.
Even as a fundamental theme,
    this question is very challenging and by no means resolved
    yet.
To derive the high order virial coefficients of strongly correlated
    fermions in the unitary limit,
    the involved quantum statistical fluctuation/correlation and
    detailed dynamical effects must be clarified, which is the novel
    systematic requirement for the sound theoretical efforts.
With the aim of calibrating the universal
    virial coefficients,
    we strive to examine the spontaneously quantum level shift contribution
    on the entropy (counting the microscopic states) in a dynamically and
    thermodynamically self-consistent way.

In this work, the dynamical and statistical
    correlation analysis explicitly demonstrates that the dense and hot
    thermodynamics at unitarity obey the virial theorem for the ideal
    non-interacting gas, i.e., $P =2/3E/V$\refr{Ho2004,Thomas2005}.
Meanwhile, the calculated virial
    coefficients at unitarity are found to be proportional to those for the ideal Fermi gas with
    integer ratios through an universal \textit{general term formula}.

In the strongly interacting system, the dynamical effects compete
with the non-linear  fluctuation/correlations. The second virial
coefficient is found to be
    vanishing due to the complicated
    correlations.
From the viewpoint of bulk properties,
    is the dilute unitary Fermi gas behavior much more like the classical ideal Boyle gas in the weak degenerate Boltzmann regime?

The present paper is organized as follows. In Sec.\ref{section2},
   the statistical method with effective field theory formalism is presented.
The approach is promoted by the phase separation-instability
discussions for a compact environment containing a competitive
Coulomb frustration element. In this work, we will take into account
    the novel non-Gaussian fluctuation effects on the strongly interacting fermions thermodynamics.
The thermodynamical consistency, virial
    theorem discussion and the high order virial coefficient
calculations of the unitary Fermi system are presented in
    Sec.\ref{section3}.
The numerical results are also presented in this section in order to
allow comparison with previous investigations. In
Sec.\ref{section4}, the scaling property of equation of state and
the  virial expansion in the
    unitary regime are further analyzed.
Summarizing remarks are made in Sec.\ref{conclusion}.

The calculations are performed in terms of the universal
four-fermions contact interaction formalism. The natural units
$k_B=\hbar=1$ are used throughout the paper.

\section{Statistical dynamics with non-Gaussian correlations}\label{section2}

The strongly interacting matter offers a plausible
    perspective in looking for the general statistical
    field theory methods.
In the presence of a medium, the unitary topic
    becomes very challenging because the system is strongly correlated and has no small parameter applicable for
    any controlled perturbative calculations.

Essentially, the strong correlation effects are highly non-linear
and
    the pronounced turbulent features will appear in the unitary
    system.
The conventional mean-field theory or loop diagram ring and
    ladder resummation perturbative techniques cannot be employed for the unitary fermions.
Consequently, more theoretical attempts are urgently needed to
understand the
    detailed dynamical role of the strongly interacting fermions
    thermodynamics.
Although there are tremendous updating efforts,
    a soundly exact theory concerning the behavior at unitarity is
    still not available.

With the goal of calculating the high order virial
    coefficients, the procedure developed can allow the systematic
rearrangements of the individual expansions
    while avoiding the theoretical double countings.
The thermodynamical expressions obtained are
    in parallel with those with the linear bare contact interaction formalism.
Therefore, the method developed will be referred to as the
    \textit{quasi-Gaussian/quasi-linear approximation} in order
to indicate the difference from and similarity to lowest order
mean-field theory.

\subsection{In-medium effective action}
The in-medium behavior associated with the many-body characteristic
is the key, while a non-perturbative approach is crucial. In
many-body theory, an established fundamental perspective is that
particles can change their spectrum properties in a dense and hot
strongly correlated medium. These changes will be reflected in the
mass shifts and/or in the development of excited complex spectral
property modifications. Considering the counteracting influences of
the surrounding medium and spontaneous single-particle spectrum
modification due to the off-shell dispersive effects, a
medium-scaling functional has been proposed\refr{chen2007}
    \bea\label{Hamiltonian} \tilde{H}=&&-\int d^3x
    \psi_\alpha ^*(x) (\frac{\nabla ^2}{2m}-\mu_{r\alpha}[n,T])\psi _\alpha (x)\no\\
    &&+\frac{U_{\mbox{eff}}^*[n,T]}2 \int d^3x\psi_\alpha ^*(x)\psi^*_\beta (x)
    \psi_\beta(x)\psi_\alpha (x).
    \eea
In \eq{Hamiltonian}, $\alpha,\beta=\uparrow, \downarrow$
    represent the (hyperfine-)spin projection Ising-variable.
The effective Hamiltonian is the same as the original Bethe-Peierls
    zero-range contact interaction version,
    except that the bare coupling constant
    $U_0={4 \pi a}/{m}$ is substituted by an effective medium-scaling
    functional $U_{\mbox{eff}}^*[n,T] $.
With the bare potential $U_0$ and corresponding vanishing
    $\mu_{r\alpha}$ in the vacuum limit $n\rightarrow 0$,
    the Hamiltonian $\tilde{H}$ reduces to the original version possessing a global $U(1)$ or $Z_2$ gauge symmetry.

Due to the explicit medium dependence of $U_{\mbox{eff}}^*$ in
\eq{Hamiltonian}, we introduce
    the additional counterterm $\delta {\cal H}\propto \mu_r[n,T]$,
which is enforced by the fundamental thermodynamical Hugenholtz-van
    Hove (HvH) theorem\refr{Brown2002,Hugenholtz}. Without loss of generality, we take care of the fully
    symmetric scenario with $\mu_{r\alpha}=\mu_r$.
The complementary  $\delta {\cal H}$ implies that the
    correlation effects on the single particle energy
    spectrum are further taken into account as an effective single-body
    potential or spontaneously generated  binding energy in the spirit of density functional
    theory\refr{Kohn1965}.
On average, the thermodynamic vacuum will have been
    ``shifted" by $\mu_r N$  non-perturbatively.

As remarked in Ref.\refr{chen2007,chen2006}, the
    central ingredient of this non-perturbative procedure
    is the way of deriving the correlating coupling functional $U_{\mbox{eff}}^*$.
Essentially,
    its constitution is beyond the bare Bethe-Peierls
    dynamics itself;
    i.e, the derivation of  $U_{\mbox{eff}}^*$ must be based on a more
    underlying physical law\refr{Brown2002}.
The functional $U_{\mbox{eff}}^*$ should reasonably encode the
    non-perturbative counteracting effects of the surrounding environment.

In thermodynamics, the surrounding environment plays
    a counteracting frustration role according to Le Chatelier's
    stability principle.
This  general principle accounts for the
    environment preventing an instantaneous departure from equilibrium with an
    alternating minus function\refr{Landau,chen2007},
    which is consistent with the second
    law of thermodynamics.
We find that the frustrating correlation effects of the surrounding
environment
    can be realized via a twisted composite rearrangement matrix-vertex\refr{chen2007}
\bea\label{potential}
    U_{\mbox{eff}}^*&&=\frac{U_0}{1-(m_D^2/2)U_0},\\&&\downarrow\no\\
       a_{\mbox{eff}}&&= \frac{a}{1-(2\pi
m_D^2/m)a}.\eea
 In order to making the analytical
expressions as concise
    as possible,
    the inverse scattering
    length notation  employed, $a_{\mbox{eff}}$, is defined in terms of $U_{\mbox{eff}}^* \equiv4\pi a_{\mbox{eff}}/m$.

The reader will have noticed that there is an
    alternating ``negative" sign difference in the denominator of \eq{potential} compared with the loop ring and ladder resummation perturbative techniques;
    i.e., the medium-scaling potential \eq{potential} appears as an instantaneously anti-screening
     formalism.
This specific {minus} sign leads to quite
    different physical motivations and calculational details.
The non-linear screening formalism makes it
    possible for us to incorporate the intermediate off-shell effects in an analytical way with 4-momentum independent
    algebra equations, i.e., instead of numerically solving the
    various coupled integral equations for the multi-points correlation
    Green functions.
This refreshing attempt is motivated by the particular conformal
    analogy and/or discursion of the zero-range unitary Fermi problem with
    the universal instantaneous Coulomb correlation discussions in a
    compact confinement environment\refr{chen2005,chen2003} based
    on the relativistic continuum Dirac field theory formalism \refr{walecka1974,walecka1976},
    where involves the complex oscillatory instabilities with short
     range and long-range
    force's competitions.

By analogy to the generalized Dyson-Schwinger calculations with
    finite temperature Green function theory\refr{chen2007,chen2006}, but more conveniently,
    the Debye mass parameter $m_D^2 $ in \eq{potential} can be
    alternatively given by the generalized Ward-Identity
\bea
    m_D^2=\(\frac{\partial n}{\partial \mu^*}\)_T=\frac{2}{T\lambda^3}f_{1/2}(z')\equiv
    2\chi',
\eea where $\mu^*$ is the effective chemical potential, the
collective implicit variable defined below. At $T=0$, it reduces to
    the familiar $m_D^2=k_f
    m/\pi^2=2 N(\epsilon _f)$, where $N(\epsilon _f)$ is the unperturbated density of
    states on the Fermi surface for one component
    fermions\refr{Pethick2002}.
$m_D^2$ characterizes the fluctuation physics because it is
    related to the well-known Pauli paramagnetic spin-spin or particle number susceptibility $\chi=\frac{1}{2}\({\partial
    n}/{\partial \mu}\)_T$ (with an additional factor $\frac{1}{2}$ due to the degenerate degrees of freedom of the two-components symmetric system) according
to \bea
    \({\partial n}/{\partial \mu}\)_T=\({\partial
    n}/{\partial \mu^*}\)_T\({\partial \mu^*}/{\partial \mu}\)_T.
\eea

\subsection{Grand thermodynamical potential with quasi-Gaussian approximation}
From the general Lagrange multiplier viewpoint,
    what we will perform is an evaluation of the relative minimum $\tilde{\Omega} (T, \tilde{\mu})$ of
    the shifted $\langle\tilde{0}|H-(\mu -\mu _r) N|\tilde{0}\rangle$
    instead of directly evaluating the challenging absolute minimum
    $\Omega (T, \mu)$ of the grand thermodynamic potential
    $\langle0|H-\mu N|0\rangle$\refr{chen2007,chen2006}.
With auxiliary physical constraints, the realistic $\Omega (T, \mu)$
    incorporating the thermodynamical vacuum fluctuation and correlation effects is indirectly derived from the
    former for the given chemical potential $\mu$.

Firstly, the shifted relative minimum is evaluated by freezing the
medium-dependent
    interaction potential with the conventional condensation
    formalism\refr{chen2007,chen2006}
\bea\label{pressure} \frac{{\tilde \Omega}(T,\mu^*)}{V}&&=-P+\mu_r n \no\\
    &&=-\frac{\pi a_{\mbox{eff}}}{m}n^2-2 T \int _k \ln(1+e^{-\beta (\frac{{\bf
    k}^2}{2m }-\mu^*)}),\eea
with $\int _k=\int
    d^3{\bf k}/(2\pi)^3$ being the momentum integral.
The constraining self-consistent equation for the single particle
Green function gives the definition of the effective chemical
potential $\mu^*$
    according to\bea\label{chemical}\tilde{\mu}=\mu-\mu_r=\frac{2\pi
    a_{\mbox{eff}}}{m}n+\mu^*.\eea
The total number density $n=n_\uparrow+n_\downarrow=2n_\uparrow$ is
given by \bea \label{density} 2\int_k
    f_k\equiv n(T,\mu^*),
\eea with the defined  quasi-particle Fermi-Dirac
    distribution functions
\bea\label{dirac} f_k&&=\frac{1}{z'^{-1}e^{\beta \frac{{\bf
    k}^2}{2m}}+1},~~~
    z'=e^{\beta\mu^*}.\eea

Different from the multiplier chemical potential $\mu$,
    the dynamically collective variable $\mu^*$ characterizes the additional correlation
    effects.
With the chemical potential correction,
    the {\em effective fugacity} $z'$ is analogous to the
    conventional one $z=e^{\beta\mu}$.
Employing the quasi-particle Fermi-Dirac distribution functions
    \eq{dirac}, the thermodynamical formulae can manifest the
    standard Fermi integrals $f_{j}(z')$ with
    $j=-\frac{1}{2},\frac{1}{2},\cdots$.

Secondly, the remaining task is to determine the correction term
with physical constraints, from which
    the realistic grand thermodynamical potential will be uniquely determined.
The shift terms $\propto \mu_r$ in the
    pressure and chemical potential are canceled out by each other in
    the Helmholtz free energy density $f=F/V=-P+\mu n$.
However, the analytical expression for the relative shift strength
$\mu_r$
    can be indirectly derived from the physical free
    energy density by relaxing the {\em medium dependence} of
    interaction potential.

Invoking the thermodynamical
    relations, one can have
\bea\label{pressure-check} P&&=n^2\(\frac{\partial \(f/n\)}{\partial
    n}\)_T,~~~\\\label{pressure-check1}
    \mu &&=\(\frac{\partial f}{\partial n}\)_T.
     \eea
Comparing the results derived from \eq{pressure-check} and
\eq{pressure-check1} with
    \eq{pressure} or \eq{chemical},
    one has the definite analytical expression of the relative
    shift strength
\bea \mu_r[n,T]={\cal C}(T,\mu^*)
    \(\frac{2\pi a_{\mbox{eff}}}{m}\)^2n^2.
\eea The employed correlation factor ${\cal C}$ is defined by
    \bea
    {\cal C} (T,\mu^*)&&\equiv
    m_D\(\frac{\partial m_D}{\partial n}\)_T=\frac{f_{-1/2}[z']}{2 Tf_{1/2}[z']}.
\eea

$m_D^2$ and ${\cal C} (T,\mu^*)$ are related to the high order
    density/spin susceptibilities; i.e., they have the crystal clear physical connotations.
With these two physical variables, the integrated grand partition
function \eq{pressure} or scaling equation of state and chemical
potential
    \eq{chemical} are reduced to
\num\bea\label{final} P&&=P_{\mbox{ideal}}(T,\mu^*)+\frac{\pi
    a_{\mbox{eff}}}{m}n^2 +{\cal C}
    \(\frac{2\pi a_{\mbox{eff}}}{m}\)^2n^3,\\
\label{final2} \mu&&=\mu^*+\frac{2\pi a_{\mbox{eff}}}{m}n +{\cal C}
    \(\frac{2\pi a_{\mbox{eff}}}{m}\)^2n^2,
\eea\enum with \bea
P_{\mbox{ideal}}\equiv\frac{2T}{\lambda^3}f_{5/2}(z'),\eea which is
similar to the ideal Fermi gas
    but with the effective chemical potential as the collective variable.

In \eq{final} and \eq{final2},
    the first two terms appear as those obtained with the
    canonical Gaussian integral formalism via the frozen
    interaction; i.e., they have exactly the same structure of mean-field theory with linear-like interaction.
The last curious shift term in \eq{final}
   non-Gaussianly proportional to the cubic particle number density, $\propto n^3$, characterizes
    the high order contributions beyond the former.
This spontaneously generated term is picked up in a
    thermodynamical way by resuming the medium dependence of the interaction matrix.
Physically, this rearrangement effect considerably
    shifts the chemical potential of the particle distribution function as displayed by \eq{final2}.

The correlation contributions manifested by the
    factor ${\cal C}$  are explicitly combined with the
    dynamical high order ones.

It is of crucial importance that the low and high order
    contributions are mixed with each other through a collective variable, $\mu^*$, effective chemical
    potential.
One can see that the grand thermodynamical potential $\Omega (T,\mu
    )$ is not the naive polynomial expanded according to the bare vacuum
    interaction strength $U_0$. The dependence of the grand thermodynamical potential $\Omega$
    on the collective correlation variable $\mu ^*$
    can be numerically eliminated in favor of the realistic physical chemical potential $\mu$.
It is physically pretty satisfactory that the equations
    \eq{final} and \eq{final2} include the
    highly non-linear or turbulent correlation
    contributions.
In the weak coupling limit,
    the equations readily reduce to those in terms of the lowest order mean-field theory,
    where the non-linear fluctuation/correlation contributions disappear.

\section{Thermodynamics with quantum rearranging correlations}\label{section3}

We now examine the thermodynamical quantities with the
    self-consistent equations or the grand partition function \eq{final} with \eq{final2}.
The scenario we present is quite general although we have restricted
ourselves to the
    low energy long-wavelength thermodynamics of the ideal fully symmetric system.
For instance, the formalism and expressions can be easily extended
to the asymmetric fermions phase separation analysis with unequal
    populations\refr{chen2008}.

\subsection{Thermodynamical consistencies, entropy and energy densities}

As discussed above, the non-Gaussian correlation
    effects considerably complicate the one-one Legendre corresponding
    relation of particle number $N$ and
    the multiplier chemical potential $\mu$ in the strongly interacting systems.
Due to the explicit high order corrections for the physical chemical
    potential, the calculations require additional great care; i.e., one
    must check if the thermodynamical relations are exactly ensured
    before making further discussions.

For instance, an obvious check is to verify that the partial
    derivative of $P$ according to $\mu $ gives again the particle number density $n=N/V$ \eq{density}
\bea\label{chemical-check} n=\(\frac{\partial P}{\partial \mu
}\)_T=\(\frac{\partial
    P}{\partial \mu^* }\)_T\(\frac{\partial \mu^*}{\partial \mu }\)_T=\frac{2}{ \lambda^3}
    f_{3/2} (z').
\eea

\eq{chemical-check} is one of the stringent thermodynamic
    conditions which are automatically satisfied by our non-perturbative
    approach.
Therefore, the grand partition function \eq{final} can
    give the exact virial coefficients.

Furthermore, the energy density and heat capacity can be given by
    the dimensionless virial coefficients in the dilute Boltzmann
    regime.
This provides a novel requirement for the effective field theory.
The complication of the temperature dependent interaction provides
    additional task to check the thermodynamic consistency characterized
    by the one-one Legendre correspondence relation between $T$ and entropy $S$.

According to the thermodynamical relations for $s=S/V$ and
    $\epsilon=E/V$
\num\bea
 s&&=\(\frac{\partial P}{\partial T}\)_\mu,\\
    \epsilon&&=-T^2\frac{\partial }{\partial T}\(\frac{-P+\mu n}{T}\)_n,
\eea\enum
 and using partial derivative formulae such as
\num\bea \(\frac{\partial
    \mu^*}{\partial T}\)_{\mu}&&\(\frac{\partial T}{\partial
    \mu}\)_{\mu^*}\(\frac{\partial \mu}{\partial \mu^*}\)_{T}=-1,\\
    \(\frac{\partial {m_D^2}}{\partial T}\)_n&&=\(\frac{\partial m_D^2}{\partial
    T}\)_{\mu^*}+\(\frac{\partial {m_D^2}}{\partial \mu^*}\)_T\(\frac{\partial {\mu^*}}{\partial
    T}\)_n,
\eea\enum the integrated entropy and energy densities read
\num\bea
\label{entropy} s&&=s_{\mbox{ideal}}+{\cal D} \(\frac{2\pi
    a_{\mbox{eff}}}{m}\)^2n^2,\\\label{energy}
\epsilon &&=2\int _k
    \frac{k^2}{2m}f_k+\frac{\pi a_{\mbox{eff}}}{m}n^2+T {\cal D}\(\frac{2\pi
    a_{\mbox{eff}}}{m}\)^2n^2.
\eea\enum In the above equations, we have employed the notation
 \bea
    s_{\mbox{ideal}}\equiv-2\int _k \[f_k\ln f_k+(1-f_k)\ln (1-f_k)\],
\eea with that for the correlation factor \bea\label{mixing} {\cal
D}(T,\mu^*)=\frac{\(\frac{\partial ^2n}{\partial
{\mu^*}^2}\)_T\(\frac{\partial
    n}{\partial T}\)_{\mu ^*}-\frac{\partial ^2n}{\partial T\partial {\mu ^*}}\(\frac{\partial n}{\partial
    {\mu^*}}\)_T}{2\(\frac{\partial n}{\partial \mu^*}\)_T}.
\eea

The first term in entropy density \eq{entropy} and first two terms
    in energy density \eq{energy} constitute the conformal formalism of ideal
    Fermi gas or mean-field theory but with the effective chemical potential.
Significantly, there is an additional (last) spontaneously generated
    term $\propto {\cal D}(T,\mu^*)$ resulting from the quantum level rearrangement.
These extra terms are the explicit dynamical high order
    contributions and of non-Gaussian characteristic.
As explicitly indicated by \eq{mixing}, the high order effects are
mixed with the fluctuations through the temperature and
    particle number density susceptibilities.
These contributions are vanishing in the zero-temperature
   conditions.
Therefore, they do not affect the previous zero-temperature
universality property discussions\refr{chen2007,chen2006,chen2005}.
Furthermore, they are canceled out by each other exactly in the
    physical Helmholtz free energy density $f=\epsilon-Ts$ and consequently can be easily neglected.
The present recipe rectifies the drawback.

\begin{figure}[h!]
        \psfig{file=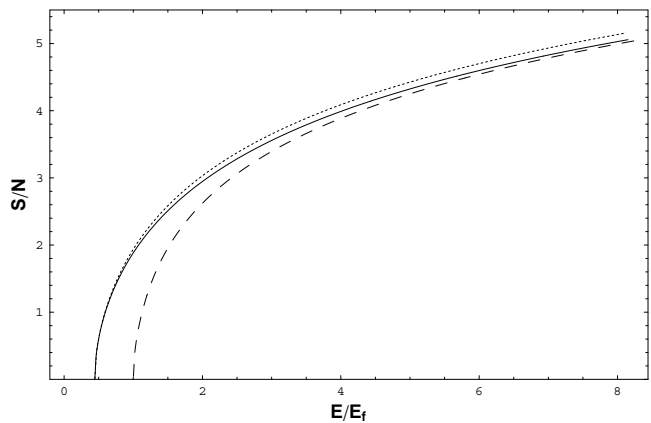,width=8.5cm,angle=-0}
        \caption{Entropy per particle versus rescaled energy at unitarity\refr{Luo2007}. The solid curve is for the
    theoretical value including the spontaneously generated contribution while the dotted one is for
the mean-field theory result. The below dashed curve is that for the
free Fermi gas.}\label{fig1}
\end{figure}

With the analytical expressions, we have indicated the entropy per
particle versus the rescaled energy in Fig.\ref{fig1}. One can see
that although the high order contributions can be almost neglected
in the low temperature regime, they play a significant role with the
increase of temperature (energy density). In comparison to that for
the mean-field theory without the non-Gaussian correlation
contribution, the convex behavior of the entropy curve more closely
approaches the experimental result or numerical
simulations\refr{Luo2007}. For comparison, we have also given the
curve for the ideal free Fermi gas.

\subsection{Virial theorem and universal coefficient $\xi (T)$ at finite temperature}

In order to justify the validity of our quasi-Gaussian
approximation, the virial theorem for the dense and hot unitary
Fermi gas will be strictly re-examined and proved.

The non-Gaussian correction terms appear in the pressure and energy
density, simultaneously. At unitarity, the general analytical
expressions for the pressure, \eq{final} and energy density,
\eq{energy}, are reduced to \bea \label{ratio}
\epsilon&&=\frac{3T}{\lambda ^3}\(
    f_{5/2}(z')-\frac{f^2_{3/2}(z')}{2f_{1/2}(z')}+\frac{f^3_{3/2}(z')f_{-1/2}(z')}{2f^3_{1/2}(z')}\)\no\\
    &&=\frac{3}{2} P.
\eea \eq{ratio} explicitly demonstrates that the unitary Fermi gas
thermodynamics
    obeys the virial theorem of ideal gas\refr{Ho2004,Thomas2005}.
What we want to emphasize is that the crucial high order
    non-Gaussian fluctuation terms ensure the virial theorem and the
    zero-temperature HvH theorem.

The energy density manifests
    the scaling property as displayed by \eq{ratio}.
The universal coefficient $\xi(T)$, i.e, the
    ratio of the unitary Fermi gas energy density to that of the ideal
    ones, can be given as a function of $z'$
\bea
    \xi (T)=1-\frac{f_{3/2}^2(z')}{2f_{1/2}(z')f_{5/2}(z')}+\frac{f_{3/2}^3(z')f_{-1/2}(z')}{2f_{1/2}^3(z')f_{5/2}(z')}.
\eea The effective fugacity $z'$ is uniquely determined by the
    particle number density according to \eq{density} or
    \eq{chemical-check} for fixed temperature $T$.

\begin{figure}[h!]
        \psfig{file=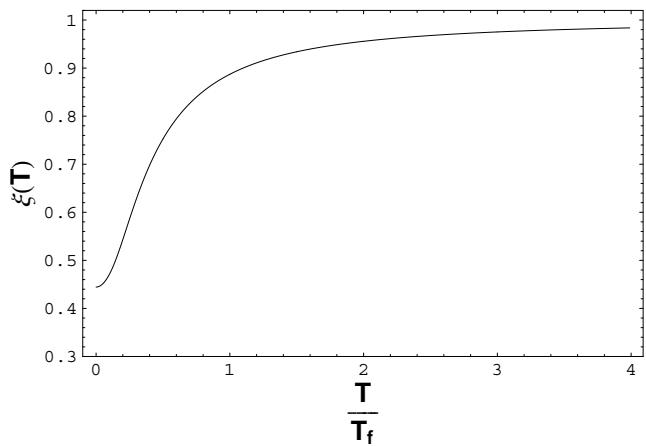,width=8.5cm,angle=-0}
        \caption{The universal coefficient $\xi(T) $ versus rescaled
temperature.}\label{fig2}
\end{figure}

As shown in Fig.\ref{fig2}, the universal coefficient approaches the
    expected saturation value $\xi=1$ in the Boltzmann limit.
It is worthy noting that in the low temperature strong degenerate
    regime with $T_f/T\gg 1$,
    the effective fugacity can be explicitly indicated as $z'=e^{T_f/T}$,
    where $T_f$ is Fermi characteristic temperature.
As stressed in \refr{chen2007,chen2006}, the zero-temperature
    result $\xi(0)=\frac{4}{9}$ is  consistent with the existing theoretical calculations \refr{physics/0303094,Steele} and experimental measurements\refr{Luo2007,Partridge2006,Stewart2006}.

\begin{figure}[h!]
\psfig{file=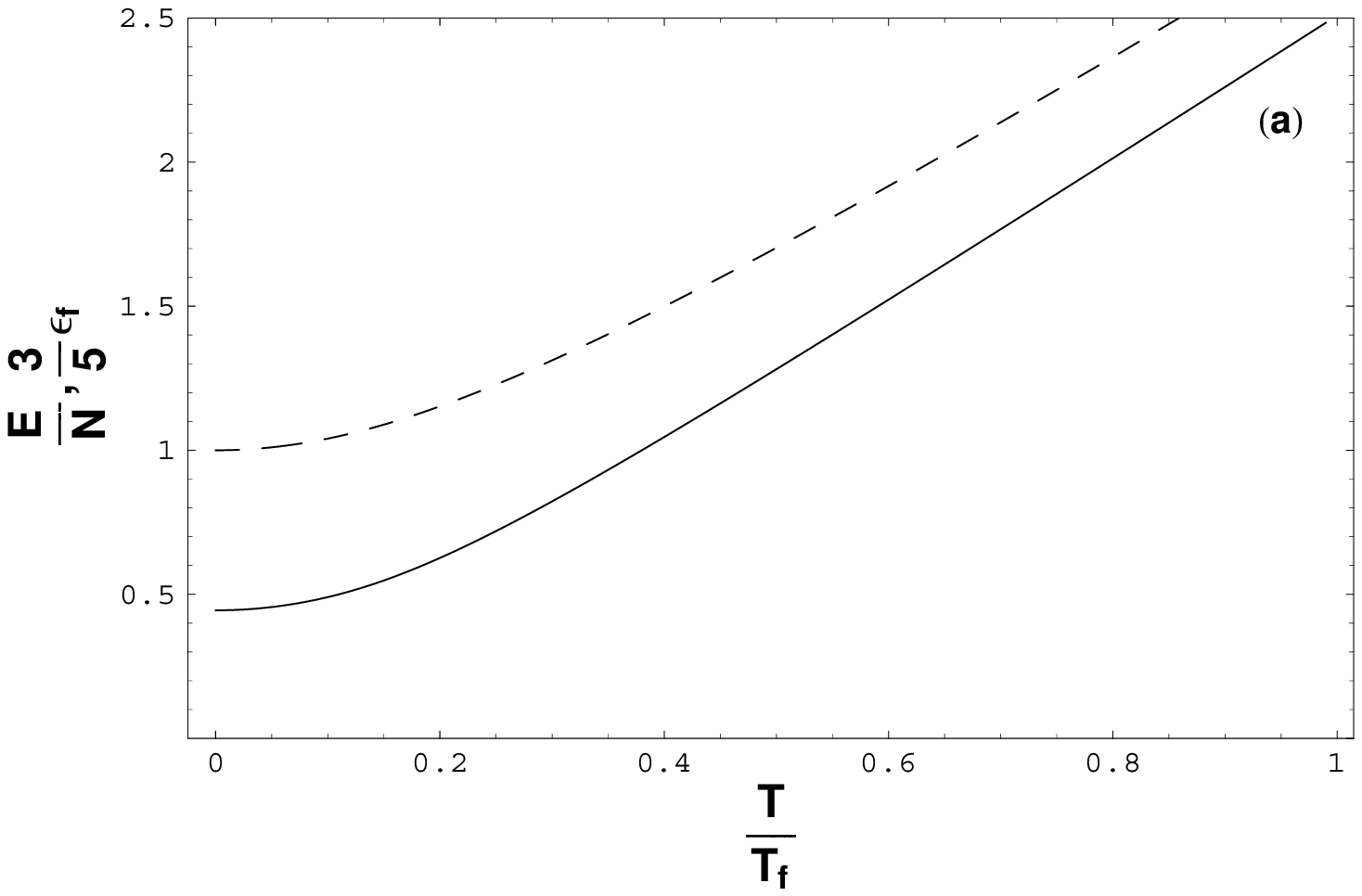,width=8.5cm,angle=-0}\\
\psfig{file=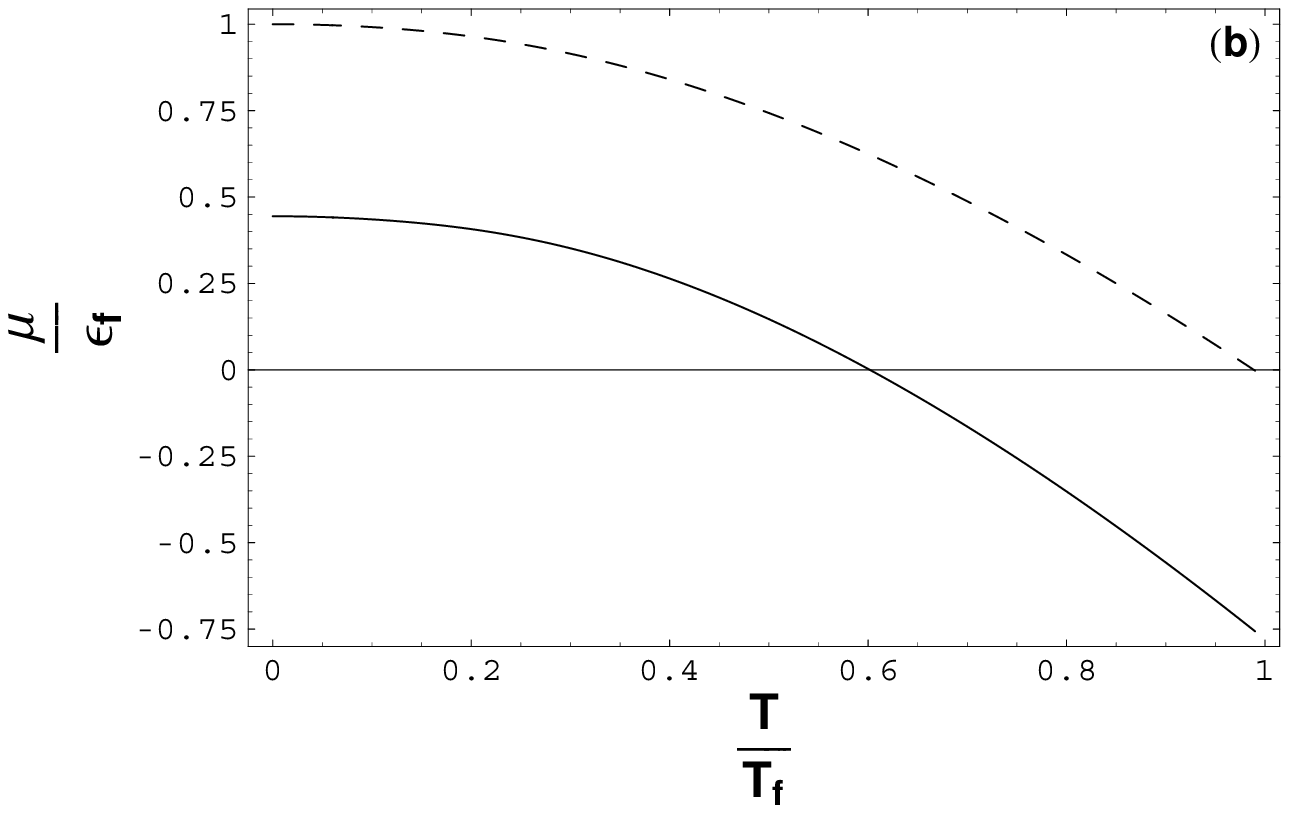,width=8.5cm,angle=-0}
        \caption{a),  Energy per particle versus rescaled temperature; b),
the physical chemical potential. The solid curves are for the
    results of unitary Fermi gas. The dashed ones are those for the
free Fermi gas.}\label{fig3}
\end{figure}

In Fig.\ref{fig3}, we have presented the finite temperature
thermodynamical numerical results. They are
    reasonably consistent with previous theoretical investigations\refr{Bulgac2005-1,Burovski2006}.

\subsection{Virial coefficients in the Boltzmann regime}

To shed light on the power of the quasi-Gaussian approximation,
    we will consider the virial expansion for the dilute unitary Fermi system.

From the general statistical mechanics viewpoint, the virial
equation of state is a series of pressure expanded in density. For
the low density weak degenerate scenario $n\lambda^3<1$, the
equation
    of state can be expanded according to \bea\label{varial1}
    \frac{P}{nT}=\sum _{l=1}^{\infty }a_l(T)
    \(\frac{n\lambda^3}{g}\)^{l-1}.
\eea The $a_l(T)$ is  the $l$-order Clausius virial coefficient with
$g$ being the spin degenerate factor ($g=2$ in this work for the
ideal symmetric scenario).

For the purpose of characterizing the complex correlation
    effects, it is worthy noting that the expansion is according to the
    {effective fugacity} $z'=e^{\beta \mu^*}$ instead of directly
    according to $z=e^{\beta \mu }$.
Meanwhile, in order to avoid the
    confusion with the conventional expansion in terms of $z$, the
    primes on $b$ and $c$ in the  expansion identities below have been
    explicitly indicated.
In the weak degenerate high temperature
    Boltzmann regime and by using such as\refr{Huang,Pathria}
\bea
    f_{3/2}(z')=z'-\frac{z'^2}{2^{3/2}}+\frac{z'^3}{3^{3/2}}-\frac{z'^4}{4^{3/2}}+\cdots,
\eea the pressure $P$ and particle number density $n$ are
    the functions of small $z'$ and can be expanded as
\num\bea \label{varial2} \frac{P}{T}&&=\frac{2}{\lambda^3}\sum
_{l=1}^\infty
    b_l'z'^l,~~~\\\label{varial3}
    n&&=\frac{2}{\lambda^3}\sum _{l=1}^\infty c_l'z'^l.
\eea\enum

By comparing \eq{varial1} with \eq{varial2} and \eq{varial3}, the
relations between $a_l'$, $b_l'$ and $c_l'$ are derived as
\num\bea
    a_1&&=\frac{b_1'}{c_1'},\\
    a_2&&=b_2' -c_2',\\ a_3&&=b_3' - 2b_2'c_2' + 2c_2^{'2} - c_3',\\
    a_4&&=b_4' - 3b_3'c_2' + 5b_2c_2^{'2} - 5c_2^{'3} - 2b_2'c_3' + 5c_2'c_3'
    -c_4',\no\\ &&\cdots .
\eea\enum It is worthy noting that the explicit results
$b_1'=c_1'=1$ have been used to reduce the relations.

With the exact grand thermodynamical
    potential \eq{final} and corresponding particle number density expression \eq{chemical-check},
    one can immediately obtain the dimensionless virial coefficients
    at unitarity $|a|=\infty$
\num\bea
 a_1&&=1,\\
    a_2&&=0,\\
 a_3&&=-\frac{1}{4}+\frac{4}{9\sqrt{3}}=-2 a^{(0)}_3,\\
    a_4&&=-\frac{15}{32}-\frac{25}{32\sqrt{2}}+\frac{5}{2\sqrt{6}}=-5
    a^{(0)}_4,\\
&&\cdots\no\\\label{general}
 a_i&&=-\frac{(i+1)(i-2)}2a_i^{(0)}, \\
&&\cdots.\no \eea\enum
 The ratios of the virial
coefficients for the Unitary Fermi gas to their
    counterparts $a^{(0)}_l$s of the ideal non-interacting Fermi gas are {integers} through a general term formula \eq{general}.

\subsection{The second virial coefficient with non-Gaussian
fluctuations} From the quantum degenerate viewpoint and as in the
classical van der Waals equation of state, the second virial
coefficient is usually considered to be ``most important". Compared
with $a_2^{(0)}=\pm
    \frac{1}{4 \sqrt{2}}$ of the ideal
    quantum gas with the effective quantum ``attraction" ($-$) for
    bosons and ``repulsion" ($+$) for fermions, the second virial coefficient of unitary Fermi gas is found to be vanishing.

The vanishing $a_2=0$ is different from
    what was reported in the literature.
In \refrs{Ho2004,Ho2004-1}, the  simple application of the
conventional quantum cluster expansion
technique\refr{Pathria,Huang,Beth} gives $a_2/a_2^{(0)}\sim -3$
(note the \textit{minus} sign). The differences between those
reported in the literature and ours manifest on two respects:
quantitatively, the magnitude of the second virial coefficient
obtained in our approach is
     smaller than $a_2^{(0)}$; qualitatively, it is vanishing.

What physical reasons lead to the second virial
    coefficient $a_2=0$ with the explicit difference?
In applying the perturbative quantum cluster expansion for
calculating the second virial coefficient, the dilute condition
$n|f_0|^3 \ll 1$ is assumed in addition to $n\lambda^3\ll 1$.
Although the divergent scattering amplitude should drop out in the
final thermodynamical quantities, the dilute condition $n|f_0|^3\ll
1$ is not automatically satisfied.
In the language of resummation
    technique,
    the effects resulting from the infinite dynamical high orders
    with various irreducible expansion diagrams also contribute to the second virial coefficient.

In the strongly correlated medium with $n|f_0|^3\gg 1$, especially
at unitarity, the non-linear fluctuation effects and
    dynamical high order contribution have mutual influence.
The microscopic attractive dynamics and repulsive quantum
    correlations are mixed with each other.
Their contributions for the second virial coefficient are found to
    be offset at unitarity; i.e, the competitive non-linear fluctuation effects make the
    expected leading order quantum correction to Boyle's law
    vanishing.

On keeping the virial expansion up to the second order for the
equation of state,
    the macroscopic thermodynamical property indicates an explicit Boyle
tendency of a classical ideal gas. Therefore, the complicated
non-Gaussian effects in the strongly correlated quantum system can
be directly validated by measuring the first-order Joule-Thomson
coefficient in the Boltzmann regime.

\section{Prospective scaling properties and virial
expansion of unitary Fermi gas}\label{section4}

One the one hand, the essential divergence nature of the bare
two-body scattering amplitude $f_0(k,a)=i /k $ implies that the weak
coupling order-by-order ``expansion" is
    not applicable with a real or virtual
    shallow bound state.
One the other hand, the scaling property with the divergent
scattering interaction strength implies that the thermodynamic
expressions can be very simple; i.e., the analytical formulae or
    virial coefficients must be symmetrically conformal to those for the ideal noninteracting Fermi gas.

Like in the classical van der Waals equation of state, the lowest
    order of perturbative quantum cluster expansion in calculating $a_2$ is actually
    on the mean-field theory level.
In performing virial expansion at unitarity, the non-linear
fluctuation effects should be reasonably well taken into account.
The difficulty is reflected by the fact that the individual low and
high order dynamical contributions cannot be separately described
from the quantum fluctuations.

The fluctuation effects on the thermodynamics quantities beyond the
mean-field theory can be described by the particle number density
susceptibility. Of course, the susceptibility itself is the final
goal. With this quasi-Gaussian approximation method, the integrated
equation of state is uniquely determined by the particle number
susceptibility $\chi$ indirectly in terms of the auxiliary  variable
$\mu^*$, correlation factors ${\cal C}$ and ${\cal
    D}$.
Through $\mu^*$, the thermodynamical quantities can be given by a
    set of highly non-linear coupled equations.
In the high temperature weak degenerate Boltzmann regime, the
    analytical virial expansion can be performed by eliminating the
    auxiliary variable.

As indicated by the coupled equations \eq{final} and \eq{final2},
   the non-Gaussian fluctuation contributions compete with the high order dynamical effects.
Their effects on the second virial coefficient are found to exactly
    cancel each other out at unitarity.
Therefore, from the
    viewpoint of virial expansion to explore the
    quantum degenerate physics, one must go beyond the calculation of $a_2$.

A general term formula connecting the virial coefficients of ideal
Fermi gas and unitary Fermi gas is found for the high order virial
coefficients. With the scaling equation of state \eq{ratio},
    the calculated virial coefficients
    manifest the expected scaling properties.
It is worthy noting that the
    {negative} sign in front of $a_4=-5a_4^{(0)}$ indicates
    a bosonization conversion tendency of the unitary Fermi gas at the {\it intermediate} BCS-BEC crossover point.

\section{Conclusion}\label{conclusion}
In summary, the three-dimensional strongly correlated BCS-BEC
    crossover thermodynamics is discussed with an auxiliary fluctuating
    mirror background shift method.
The equations presented demonstrate a mutual influence of the
    quantum statistical correlation and dynamical effects for different orders.

The goal of this work is to make an analytical attempt
    doing the challenging non-linear virial series expansion.
The obtained results manifest the scaling property of unitary Fermi
    gas; i.e., the virial coefficients are found to be proportional to
    those of ideal Fermi gas with a general formulae.

In contrast to the investigations in the literature, the second
virial coefficient of the unitary Fermi gas is found to be
vanishing. To a great extent, this unpleasant result further puzzles
us as regards reaching a deeper theoretical understanding of the
BCS-BEC crossover physics in the Boltzmann regime.

In physics, $a_2=0$ would not be unexpected because the unitary
Fermi system is \textit{in between} the dimer boson and fermion
phases. The non-linear fluctuation effects are mixed with and
compete with the dynamical effects. Their contributions to the
second virial coefficient can be exactly canceled out by each other.
By keeping up to the second order virial expansion in the weakly
degenerate regime, the bulk properties can manifest the Boyle's law
of classical ideal gas. The Joule-Thomson Coefficient measurement in
the unitary limit can clarify the difference or be used to explore
the novel non-Gaussian correlation effects, in the affirmative.

In the strongly interacting regime, especially at unitarity, the
    non-Gaussian fluctuations play an important role.
Technically, the realistic grand thermodynamical
    potential $\Omega (T,\mu) $ or equation of state is described by the two coupled
    parametric-equations \eq{final} and \eq{final2}.
Equivalent to eliminating the implicit variable, effective chemical
    potential $\mu^*$, the potential expansion is according to
    the effective fugacity $z'=e^{\beta
    \mu^*}$ instead of directly according to fugacity $z=e^{\beta \mu }$.

The virial theorem at unitarity, the general HvH theorem at zero
    temperature and the fundamental thermodynamical consistency
    relations are strictly ensured in the formulation.
This non-perturbative approach makes it possible to discuss the
    non-Gaussian fluctuation physics in the novel strongly correlated quantum system.

\acknowledgments{J.-s Chen thanks S.-p Wu for discussions  on the
quasi-linear approximation.
 This work was
supported in part by the Natural Science Foundation of China under
Grant No. 10875050, 10675052 and MOE of China
    under projects No.IRT0624, the scientific research Fund of
    Central China Normal University.
   }


\begin{thebibliography}{99}

\bibitem{Giorgini2007}
S. Giorgini, L.P. Pitaevskii, and S. Stringari,
\rmp{80}{1215}{2008}.


\bibitem{Heiselberg2000}H. Heiselberg, \pra{63}{043606}{2001}.

\bibitem{Ho2004}T.-L. Ho, \prl{92}{090402}{2004}.
\bibitem{Ho2004-1}T.-L. Ho and E.-J. Mueller,
\prl{92}{160404}{2004}.


\bibitem{Thomas2005}J.E. Thomas, J. Kinast, and A. Turlapov, \prl{95}{120402}{2005}.

\bibitem{Luo2007}L. Luo, B. Clancy, J. Joseph, J. Kinast, and J. E.
Thomas, \prl{98}{080402}{2007}.

\bibitem{Bulgac2005-1}
A. Bulgac, J.E. Drut, and P. Magierski, \prl{96}{090404}{2006}.

\bibitem{Burovski2006}E. Burovski, N. Prokof'ev, B. Svistunov, and M. Troyer, \prl{96}{160402}{2006}; \njp{8}{153}{2006}.


\bibitem{Schwenk2005}
C.J. Horowitz and  A. Schwenk, \plb{638}{153}{2006}.


\bibitem{physics/0303094}
J. Carlson,  S.Y. Chang,  V.R. Pandharipande, and K.E. Schmidt,
\prl{91}{050401}{2003}.


\bibitem{Pethick}
A. Schwenk and C.J. Pethick, \prl{95}{160401}{2005}.

\bibitem{Steele}
J.V. Steele, e-print arXiv:nucl-th/0010066.

\bibitem{Xiong2005}
H.-W. Xiong, S.-J. Liu, W.-P. Zhang, and M.-S. Zhan,
\prl{95}{120401}{2005}.


\bibitem{Hu2006}H. Hu, X.-J. Liu, and P.D. Drummond,
\epl{74}{574}{2006}.
\bibitem{Rupak2007}
G. Rupak, \prl{98}{090403}{2007}.

\bibitem{Castin2006}
Y. Castin, e-print arXiv:cond-mat/0612613.

\bibitem{chen2007}
J.-S. Chen, C.-M. Cheng, J.-R. Li, and Y.-P. Wang,
\pra{76}{033617}{2007}.

\bibitem{Brown2002}
G.E. Brown and M. Rho, \prep{363}{85}{2002}.

\bibitem{Hugenholtz}
N.M. Hugenholtz and L.van Hove, Physica (Amsterdam) {\bf 24}, 363
(1958).





\bibitem{Kohn1965}
W. Kohn and L.J. Sham, \pra{140}{1133}{1965}.


\bibitem{chen2006}
J.-S. Chen, \cpl{24}{1825}{2007} and \ctp{48}{99}{2007}.

\bibitem{Landau}
L.D. Landau and E.M. Lifshitz, \textit{Statistical Physics}
(Pergamon Press, 1980).

\bibitem{chen2005}
J.-S. Chen, J.-R. Li, and M. Jin, \plb{608}{39}{2005}; J.-S. Chen,
e-print arXiv:nucl-th/0509038.

\bibitem{chen2003}
 J.-S. Chen, J.-R. Li, and P.-F. Zhuang,
\prc{67}{068202}{2003}.




\bibitem{walecka1974}
J.D. Walecka, \ann{83}{491}{1974}.

\bibitem{walecka1976}
    B.D. Serot and J.D. Walecka, \anp{16}{1}{1986};
    \ijmpe{6}{515}{1997}.






\bibitem{Pethick2002}
C.J. Pethick and H. Smith, \textit{Bose-Einstein Condensation in
Dilute Gases} (Cambridge University Press, Cambridge, England,
2002).


\bibitem{chen2008} J.-S Chen, F. Qin, and Y.-P Wang,
 e-print arXiv:0803.4385.

\bibitem{Huang}
Kerson Huang, {\em Statistical Mechanics} (Wiley, New York, 1987).

\bibitem{Pathria}
R.K. Pathria, {\em Statistical Mechanics}, (2nd edition),
Butterworth-Heinemann, Oxford (1996).

\bibitem{Beth}
E. Beth and G.E. Uhlenbeck, Physica (Amsterdam) {\bf 4}, 915 (1937).



\bibitem{Partridge2006}G.B. Partridge, W. Li, R.I Kamar, Y.-A Liao, and R.G. Hulet,
\sci{311}{503}{2006}.

\bibitem{Stewart2006}J.T. Stewart, J.P. Gaebler, C.A. Regal, and D.S. Jin,
\prl{97}{220406}{2006}.


\end{thebibliography}
\end{document}